\documentclass[conference]{IEEEtran}

\ifCLASSINFOpdf
   \usepackage[pdftex]{graphicx}
\else
   \usepackage[dvips]{graphicx}
\fi

\usepackage[cmex10]{amsmath}
\usepackage{amssymb}
\usepackage{amsthm}
\usepackage{mathtools}
\usepackage[hidelinks]{hyperref}
\usepackage{cite}
\usepackage{soul}
\setstcolor{red}
\usepackage{lipsum}
\usepackage{xargs}
\usepackage{booktabs}
\usepackage{multirow}
\usepackage{algorithm}
\usepackage{algpseudocode}
\usepackage{tabularx}
\usepackage[normalem]{ulem}
\useunder{\uline}{\ul}{}
\usepackage[table,xcdraw,dvipsnames]{xcolor}
\usepackage{cite}
\usepackage[caption=false,font=footnotesize]{subfig}

\makeatletter
\let\old@ps@headings\ps@headings
\let\old@ps@IEEEtitlepagestyle\ps@IEEEtitlepagestyle
\def\psccfooter#1{%
    \def\ps@headings{%
        \old@ps@headings%
        \def\@oddfoot{\strut\hfill#1\hfill\strut}%
        \def\@evenfoot{\strut\hfill#1\hfill\strut}%
    }%
    \def\ps@IEEEtitlepagestyle{%
        \old@ps@IEEEtitlepagestyle%
        \def\@oddfoot{\strut\hfill#1\hfill\strut}%
        \def\@evenfoot{\strut\hfill#1\hfill\strut}%
    }%
    \ps@headings%
}
\makeatother

\def\thanksto#1{
\begingroup
\def\thefootnote{}
\footnote{
\kern -3pt
\hrule width 0.4\columnwidth height 0.2pt
\kern 5pt
#1
}
\setcounter{footnote}{0}
\endgroup
}

\begin{document}

\title{The Living Forecast: Evolving Day-Ahead Predictions into Intraday Reality}

\author{
\IEEEauthorblockN{Kutay Bölat,  Peter Palensky, Simon H. Tindemans}
\IEEEauthorblockA{Department of Electrical Sustainable Energy \\
Delft University of Technology\\
Delft, The Netherlands\\
\{K.Bolat, P.Palensky, S.H.Tindemans\}@tudelft.nl}
}

\maketitle
\global\csname @topnum\endcsname 0
\global\csname @botnum\endcsname 0
\begin{abstract}
Accurate intraday forecasts are essential for power system operations, complementing day-ahead forecasts that gradually lose relevance as new information becomes available. This paper introduces a Bayesian updating mechanism that converts fully probabilistic day-ahead forecasts into intraday forecasts without retraining or re-inference. The approach conditions the Gaussian mixture output of a conditional variational autoencoder-based forecaster on observed measurements, yielding an updated distribution for the remaining horizon that preserves its probabilistic structure. This enables consistent point, quantile, and ensemble forecasts while remaining computationally efficient and suitable for real-time applications. Experiments on household electricity consumption and photovoltaic generation datasets demonstrate that the proposed method improves forecast accuracy up to 25\% across likelihood-, sample-, quantile-, and point-based metrics. The largest gains occur in time steps with strong temporal correlation to observed data, and the use of pattern dictionary-based covariance structures further enhances performance. The results highlight a theoretically grounded framework for intraday forecasting in modern power systems.
\end{abstract}

\thanksto{\noindent This project has received funding from the European Union's Horizon 2020 research and innovation programme under the Marie Skłodowska-Curie grant agreement No 956433.}

\section{Introduction}
\begin{figure}
    \centering
    \includegraphics[width=.95\linewidth]{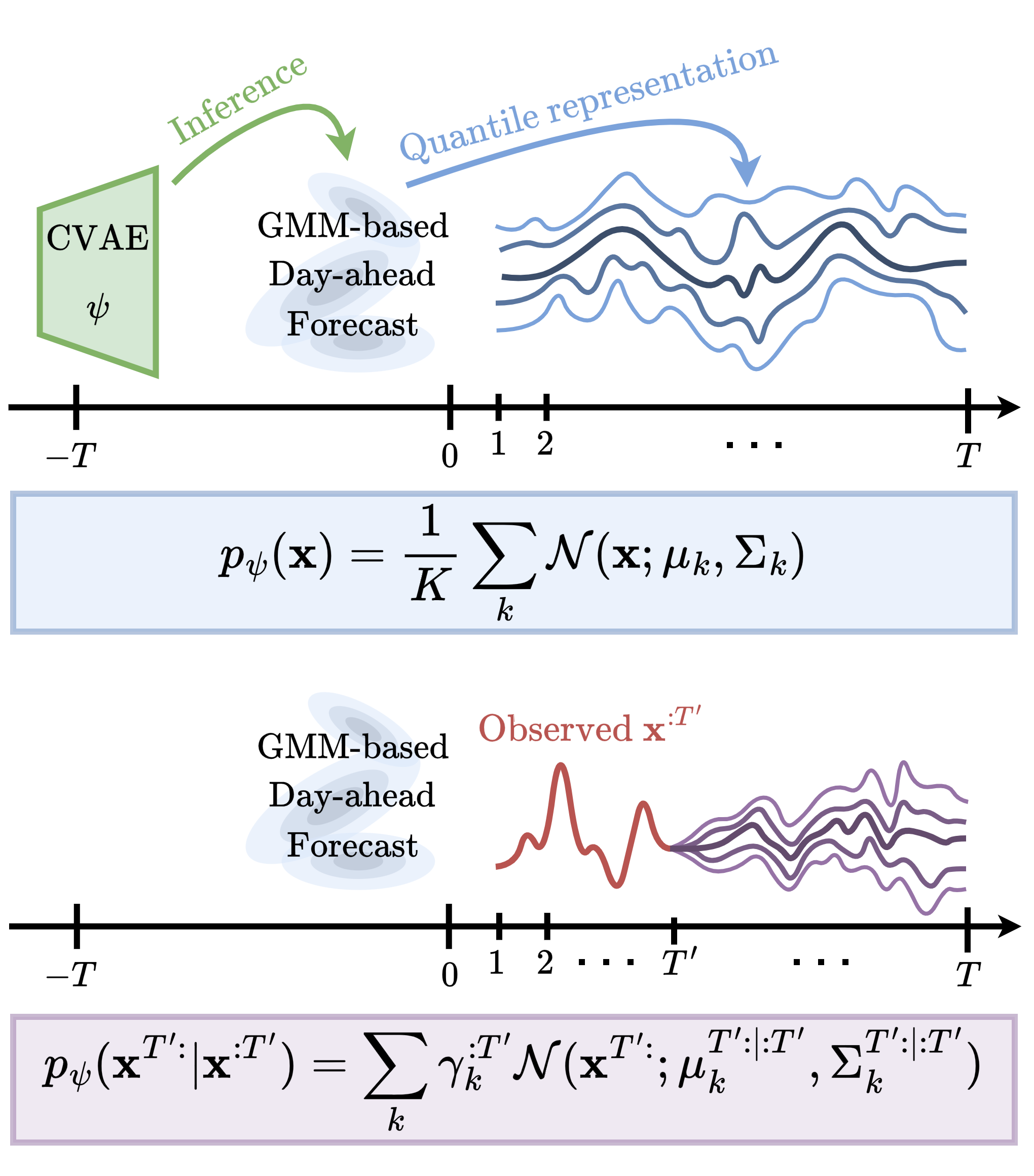}
    \caption{Overview of the proposed Bayesian intraday update mechanism. A fully probabilistic day-ahead forecast, represented as a Gaussian mixture model, is obtained from a conditional variational autoencoder (top). Given partial intraday observations, the model applies a closed-form Bayesian update to adjust the remaining-horizon distribution (bottom), enabling consistent quantile, ensemble, and point forecasts without retraining or re-inference.}
    \label{fig:update_fig}
\end{figure}

Power systems rely heavily on accurate short-term forecasts to operate efficiently under uncertainty. While day-ahead forecasts offer a global view of the upcoming day, they become progressively outdated as the day progresses and new information becomes available. Intraday forecasting addresses this gap by refining the forecast for the remaining hours of the day, leveraging partially observed measurements and updated exogenous inputs \cite{kolkmann2024modeling}. This additional forecasting layer is critical for real-time decision-making tasks such as rescheduling generation, updating market positions, and managing reserves \cite{nickelsen2025bayesian, visser2024probabilistic}. As renewable penetration and demand-side flexibility increase, the role of intraday forecasting in maintaining operational reliability becomes vital \cite{zhang2024quantitative}.

Given this operational role of intraday forecasting, the ability to quantify uncertainty is essential. While point forecasts provide a single-valued estimate of future outcomes, they fail to capture the uncertainty inherent in power system operations. Probabilistic forecasts address this by offering richer information about forecast uncertainty, enabling risk-aware decisions in scheduling, reserve allocation, and market bidding \cite{wang2023quantifying, costilla2022operating}. For example, quantile forecasts are widely used in electricity market bidding strategies, where upper and lower quantiles help balance profit maximization with the risk of imbalance penalties \cite{visser2024probabilistic, zhang2025weather}. However, these forecasts are typically produced independently for each time step, which means they provide information only about marginal distributions, not the temporal joint distribution. As a result, they do not support sampling of realistic trajectories and are limited in applications such as scenario generation or time-coupled optimization. Fully probabilistic forecasters overcome these limitations by modelling a joint probability distribution over the entire forecast horizon, conditioned on available inputs \cite{bolat2025one}. From this distribution, one can directly obtain point, quantile, and ensemble forecasts, enabling consistent and flexible use across a range of decision-making tasks. However, as the day progresses and new observations become available, these forecasts must be updated to preserve accuracy and relevance in intraday operations.

A variety of methods have been explored for intraday forecast refinement, yet each presents important limitations. One common approach is to train separate models for intraday forecasting conditioned on partial observations \cite{visser2024probabilistic}. While effective, this strategy increases training and maintenance effort and may lead to inconsistencies with the original day-ahead forecast. Adaptive schemes based on sliding-window autoregressive training update predictions using recent data, but their data requirements grow with temporal resolution, and they may struggle to preserve performance over longer horizons. Heuristic corrections, such as bias adjustment or smoothing \cite{kolkmann2024modeling}, are computationally efficient but operate primarily at the marginal level, lacking a coherent probabilistic structure. Copula-based approaches explicitly model temporal dependencies while preserving marginal distributions \cite{austnes2025probabilistic}. However, separating marginal and dependency modelling can introduce structural inconsistencies.

Recent probabilistic forecasting studies further emphasise the importance of joint uncertainty modelling in intraday power system applications. Ensemble-based mixture density networks have been applied to intraday PV forecasting to provide flexible distributional outputs \cite{doelle2023probabilistic}, and Bayesian hierarchical frameworks have been proposed for continuous intraday electricity markets \cite{nickelsen2025bayesian}. Multivariate update models for renewable infeed explicitly account for temporal correlations \cite{kolkmann2024modeling}, while hybrid approaches combining Kalman post-processing with machine learning aim to refine intraday uncertainty estimates sequentially \cite{thaker2024hybrid}. Advanced deep learning architectures integrating transformer models with Gaussian process approximations further demonstrate the potential of expressive probabilistic modelling \cite{xiong2025deep}.

Despite these advances, most existing approaches either require retraining, repeated neural network inference, or architectural modifications to incorporate new observations during the day. As a result, maintaining structural consistency with the original day-ahead joint distribution while ensuring computational scalability remains challenging.

In this work, we propose a simple and theoretically grounded approach for intraday forecast updating that directly builds on the existing day-ahead probabilistic forecast. Rather than retraining models, modifying architectures, or generating new scenarios from scratch, our method applies a Bayesian update to the original forecast distribution, conditioned on newly available intraday observations. The proposed update process is illustrated in Fig. \ref{fig:update_fig}: starting from the fully probabilistic day-ahead forecast distribution, the method conditions on the observed portion of the day to produce an updated distribution for the remaining hours. The underlying day-ahead forecast is modelled using a conditional variational autoencoder with a Gaussian mixture output, enabling rich and flexible representation of uncertainty across time. This structure enables the update to be performed directly in the distribution space, without requiring retraining or re-running the neural network. As a result, the updated forecast remains fully probabilistic, supports sampling and quantile extraction, and preserves consistency with the original model. The approach avoids retraining and additional neural network inference during intraday operation, relying instead on closed-form linear algebraic updates whose complexity scales with the number of observed time steps. This makes it suitable for integration into real-time forecasting pipelines in practical power system applications.
\section{Problem Setting}

   We consider the problem of obtaining fully probabilistic intraday forecasts by leveraging the day-ahead forecasting distribution and observed intraday measurements. Specifically, we aim to model the conditional probability density function (pdf) $p(\mathbf{x}^{T':}|\mathbf{x}^{:T'}, \mathbf{c})$ representing the intraday forecast using the day-ahead forecast pdf $p(\mathbf{x}|\mathbf{c})$. Here, $\mathbf{x}= [x_t]_{t=1}^T\in \mathbb{R}^T$ represents the daily observation vector, consisting of two intra-day partitions defined as $\mathbf{x} = [\mathbf{x}^{:T'},\mathbf{x}^{T':}]$ where $\mathbf{x}^{:T'} = [x_t]_{t=1}^{T'}$ corresponds to observations already obtained and $\mathbf{x}^{T':} = [x^{(t)}]_{t=T'+1}^{T}$ represents future intraday observations. The vector $\mathbf{c}\in\mathbb{R}^C$ denotes exogenous inputs, including historical measurements, calendar-related variables (e.g., holidays, weekdays), and external factors such as weather forecasts (e.g., temperature, wind speed).
   
   In practical terms, the objective is to ``update" the day-ahead forecast $p(\mathbf{x}|\mathbf{c})$ with new incoming measurements observed up to time $T'$ and convert it efficiently into an updated intraday forecast $p(\mathbf{x}^{T':}|\mathbf{x}^{:T'}, \mathbf{c})$. This updating process is illustrated in Fig. \ref{fig:update_fig}. Note that the fully probabilistic formulation of the forecast inherently facilitates seamless extraction of various predictive outputs, including point, ensemble, and quantile forecasts \cite{bolat2025one}.

   Formally, the intraday forecast pdf $p(\mathbf{x}^{T':}|\mathbf{x}^{:T'}, \mathbf{c})$ is obtained from the day-ahead forecast $p(\mathbf{x}|\mathbf{c})$ using Bayes' theorem:
   \begin{equation}\label{eq:bayes_update}
       p(\mathbf{x}^{T':}|\mathbf{x}^{:T'}, \mathbf{c}) = \frac{p(\mathbf{x}|\mathbf{c})}{p(\mathbf{x}^{:T'}|\mathbf{c})}.
   \end{equation}
   This relationship emerges directly from Bayesian inference, where the denominator $p(\mathbf{x}^{:T'}|\mathbf{c})$ can typically be computed by marginalizing over the original forecast distribution. However, the resulting conditional pdf does not generally remain within the same family as the original distribution, particularly when complex neural network-based distributions are employed. Consequently, our goal is to develop a day-ahead forecast distribution model whose structure is preserved under Bayesian updating, ensuring computational scalability and structural consistency in generating intraday probabilistic forecasts.
\section{Methodology}

\subsection{Day-ahead forecasting model}

We employ a conditional variational autoencoder (CVAE) \cite{sohn2015learning, bolat2025one} for modelling the fully probabilistic day-ahead forecasting model as 
\begin{equation}\label{eq:cvae_to_gmm}
\begin{split}
    p_{\psi}(\mathbf{x}|\mathbf{c}) &= \mathbb{E}_{p(\mathbf{z)}}[p_{\psi}(\mathbf{x}|\mathbf{z},\mathbf{c})] \\
    &\approx \frac{1}{K} \sum_k p_{\psi}(\mathbf{x}|\mathbf{z}_k,\mathbf{c}) =: \tilde{p}_{\psi}(\mathbf{x}|\mathbf{c})
\end{split}
\end{equation}
where $\mathbf{z}\in\mathbb{R}^L$ is the latent variable distributed as $p(\mathbf{z})$ and $ \mathbf{z}_k \overset{\sim}{\leftarrow} p(\mathbf{z})$ are the Monte Carlo samples used to approximate the intractable expectation. This results in a $K$-component mixture distribution $\tilde{p}_{\psi,\mathbf{U}}(\mathbf{x}|\mathbf{c})$, parameterized by a deep neural network $\psi$ that maps latent variables and conditions to distribution parameters.

In this work, we model each component as a multivariate normal (MVN) distribution:
\begin{equation}\label{eq:cvae}
    p_{\psi}(\mathbf{x}|\mathbf{z}_k,\mathbf{c}) = \mathcal{N}(\mathbf{x};\mu=f_\psi^\mu(\mathbf{z}_k,\mathbf{c}), \Sigma=f_\psi^\Sigma(\mathbf{z}_k,\mathbf{c}))
\end{equation}
where $f_\psi^\mu(\cdot): \mathbb{R}^{L+C} \rightarrow \mathbb{R}^{T}$ and $f_\psi^\Sigma(\cdot): \mathbb{R}^{L+C} \rightarrow \mathbb{R}^{T\times T}$ are two distinct outputs of a neural network that parameterize the mean and covariance of the MVN. This structure effectively turns the CVAE into a conditional Gaussian mixture model (GMM), with the number of components $K$ freely chosen during inference, offering flexibility and expressiveness.

A key challenge in this setup is ensuring that the covariance matrix $\Sigma$ remains positive-definite. A common solution is to restrict $\Sigma$ to be diagonal:

\begin{equation}
    f_\psi^\Sigma(\mathbf{z,c}) = \text{diag}(f_\psi^\sigma(\mathbf{z,c}))^2,
\end{equation}
where $f_\psi^\sigma(\mathbf{z,c}):\mathbb{R}^{L+C}\rightarrow\mathbb{R}_+^T$ outputs the marginal standard deviations. To overcome the limitations of diagonal covariance and enable richer representations, we employ Pattern Dictionary-based Covariance Composition (PDCC) \cite{bolat2024guide}:
\begin{equation}
    f_\psi^\Sigma(\mathbf{z,c}) = \mathbf{U}\text{diag}(f^{\tilde{\sigma}}_\psi(\mathbf{z},\mathbf{c}))^2\mathbf{U}^\top + \xi\mathbf{I},
\end{equation}
where $\mathbf{U}\in\mathbb{R}^{T\times V}$ is a learnable pattern dictionary (with $V\geq T$), $f_\psi^{\tilde{\sigma}}(\mathbf{z,c}):\mathbb{R}^{L+C}\rightarrow\mathbb{R}_+^V$ outputs auxiliary standard deviations and $\xi$ is a small constant for numerical stability. Importantly, $\mathbf{U}$ is shared across inputs and does not depend on $\mathbf{z}$ and $\mathbf{c}$, enabling the model to learn common covariance patterns across data points.

\subsection{Intraday updating mechanism} \label{subsec:intraday_update}
Once the day-ahead forecast $\tilde{p}_{\psi}(\mathbf{x}|\mathbf{c})$ is obtained, we aim to reuse it for intraday forecasting without any re-inference\footnote{We refer to re-inference as running another forward pass on the neural network.} or retraining. That is, we adapt the GMM output of the CVAE directly, as illustrated in Fig. \ref{fig:update_fig}.

Applying the Bayes’ rule in \eqref{eq:bayes_update} to the day-ahead forecast distribution in \eqref{eq:cvae_to_gmm}, the updated intraday forecast becomes:
\begin{equation}\label{eq:intraday_dist}
\begin{split}
    \tilde{p}_\psi(\mathbf{x}^{T':}|\mathbf{x}^{:T'}, &\mathbf{c})
     \\
    &= \sum_k \underbrace{\frac{\tilde{p}_{\psi}(\mathbf{x}^{:T'}|\mathbf{z}_k,\mathbf{c})}{\sum_{k'} \tilde{p}_{\psi}(\mathbf{x}^{:T'}|\mathbf{z}_{k'},\mathbf{c})}}_{\gamma_k^{:T'}} \tilde{p}_{\psi}(\mathbf{x}^{T':}|\mathbf{x}^{:T'},\mathbf{z}_k,\mathbf{c})
\end{split}
\end{equation}
which forms a $T-T'$-dimensional mixture model with weights $\gamma^{:T'}=\left[\gamma_k^{:T'}\right]_k$ lies in the (K-1)-dimensional standard simplex. For the MVN components of the day-ahead forecasting distribution $\tilde{p}_{\psi}(\mathbf{x}|\mathbf{c})$, let us define
\begin{equation*}
    \mu_k=
    \begin{bmatrix}
        \mu_k^{:T'} \\ \mu_k^{T':}
    \end{bmatrix}=f_\psi^\mu(\mathbf{z}_k,\mathbf{c})
\end{equation*}
and
\begin{equation*}
    \Sigma_k= \begin{bmatrix}
            \Sigma_k^{:T'} & \Sigma_k^{\times^\top}\\
            \Sigma_k^{\times} & \Sigma_k^{T':}
            \end{bmatrix} =f_\psi^\Sigma(\mathbf{z}_k,\mathbf{c}).
\end{equation*}
By definition, the marginal distribution up to time $T'$ becomes
\begin{equation}
    \tilde{p}_{\psi}(\mathbf{x}^{:T'}|\mathbf{z}_k,\mathbf{c}) = \mathcal{N}(\mathbf{x}_{:T'};\mu=\mu_k^{:T'}, \Sigma=\Sigma_k^{:T'}).
\end{equation}
Since each day-ahead mixture component $\tilde{p}_{\psi}(\mathbf{x}|\mathbf{z}_k,\mathbf{c})$ is an MVN, intraday components $\tilde{p}_{\psi}(\mathbf{x}_{T':}|\mathbf{x}_{:T'},\mathbf{z}_k,\mathbf{c})$ preserve the MVN structure as \cite{page1984multivariate}
\begin{equation}
    \tilde{p}_{\psi}(\mathbf{x}^{T':}|\mathbf{x}^{:T'},\mathbf{z}_k,\mathbf{c}) = \mathcal{N}(\mathbf{x}^{T':};\mu=\mu_k^{T':|:T'}, \Sigma=\Sigma_k^{T':|:T'})
\end{equation}
where
\begin{equation}\label{eq:mu_update}
    \mu_k^{T':|:T'} = \mu_k^{T':} + \Sigma_k^{\times}\Sigma_k^{:T'^{-1}}(\mathbf{x}_{:T'}-\mu_k^{:T'})
\end{equation}
and
\begin{equation}\label{eq:sigma_update}
    \Sigma_k^{T':|:T'} = \Sigma_k^{T':} - \Sigma_k^{\times}\Sigma_k^{:T'^{-1}}\Sigma_k^{\times^\top}.
\end{equation}
Intuitively, the update process can be seen as favouring the day-ahead mixture components $\tilde{p}_{\psi}(\mathbf{x}|\mathbf{z}_k,\mathbf{c})$ that best ``guessed" the observed measurements $\mathbf{x}^{:T'}$, by using $\gamma^{:T'}$ to increase the weight of their updated distributions $\tilde{p}_{\psi}(\mathbf{x}^{T':}|\mathbf{x}^{:T'},\mathbf{z}_k,\mathbf{c})$ relative to the other components for the rest of the day ($\mathbf{x}^{T':}$).

\subsubsection*{Efficient intraday sampling}

The updated distribution can be used to form an ensemble forecast $\{\hat{\mathbf{x}}^{T':}_s\}_{s=1}^S$ by sampling $S$ equiprobable traces from it. This sampling can be done in a hierarchical manner:
\begin{enumerate}
    \item Randomly select a component: ${\hat{k}_s}\overset{\sim}{\leftarrow} \text{Categorical}(\gamma^{:T'})$
    \item Calculate and store $\mu_{\hat{k}_s}^{T':|:T'}$ and $\Sigma_{\hat{k}_s}^{T':|:T'}$ using \eqref{eq:mu_update} and \eqref{eq:sigma_update} if not stored before
    \item Sample from the selected and updated component: $\hat{\mathbf{x}}^{T':}_s\overset{\sim}{\leftarrow} \mathcal{N}(\mathbf{x}^{T':};\mu_{\hat{k}_s}^{T':|:T'}, \Sigma_{\hat{k}_s}^{T':|:T'})$
\end{enumerate}
Note that the storing in the second step can significantly reduce the computational cost as \eqref{eq:mu_update} and \eqref{eq:sigma_update} are computed only once for a sampled $\hat{k}_s$ and \mbox{re-used} later if it gets sampled again.
This is particularly effective when the number of samples $S$ is large or the weights in $\gamma^{:T'}$ are sparse, i.e. most entries are near zero.

\subsubsection*{Computational considerations}
The intraday update requires only closed-form linear-algebraic operations and no additional neural network inference. For each mixture component, the dominant cost stems from inverting $\Sigma_k^{:T'} \in \mathbb{R}^{T' \times T'}$, resulting in a per-component complexity of $\mathcal{O}(T'^3)$, where $T'$ denotes the number of observed time steps. The remaining matrix multiplications scale as $\mathcal{O}((T-T')T'^2)$. Since $T'$ increases gradually during the day and remains moderate in practical settings, the update remains computationally tractable. Moreover, conditional parameters are computed once per component and cached during sampling, and sparsity in mixture weights further reduces effective run time.
\section{Experiments}

\subsection{Datasets}

\subsubsection{Gipuzkoa SM}
We employed a smart meter dataset capturing hourly electricity consumption in kWh for 25,559 residential, office, and business customers across 47 Spanish provinces from November 2014 to June 2022 \cite{quesada2024electricity}. Experiments were conducted on a subset of the dataset, consisting of records from the data-dense Gipuzkoa region and a one-year period from June 2021 to May 2022. The final dataset includes 6,830 customers with one full year of continuous data, resulting in approximately 2.5 million records (365 days per customer, $T=24$ hourly readings per day). Data preprocessing involved excluding energy-producing and consistently zero-consumption users, applying zero-preserved log-normalization with hyperparameters from \cite{bolat2024guide}, and partitioning the dataset into training, validation, and testing sets (60\%, 20\%, and 20\%, respectively). As in \cite{bolat2025one}, day-before consumption profiles and cyclically encoded (\( \sin \)-\( \cos \)) months (Jan, Feb, etc.), days of the week (Mon, Tue, etc.) were used as the exogenous inputs.

\subsubsection{Utrecht PV}
The other dataset used is a 4-year collection of household PV power production measurements in kW from 175 households in Utrecht, Netherlands \cite{visser2022open}. The 1-minute resolution measurements were aggregated to 15-minute values for this study ($T=96$), and daily profiles were extracted. The last 28 time steps (measurements after 17:00) of the daily profiles were excluded from performance assessments, as they are dominated by zeros and bias the representativeness of the metrics. Each household's measurement is normalized by its estimated AC capacity (peak output power), and the dataset is partitioned with the same ratios as the Gipuzkoa SM dataset. The exogenous inputs consisted of 24-hour weather forecasts (surface solar radiation, clear sky irradiance, medium and total cloud coverages) for Utrecht, in addition to the day-before generation profiles.

\subsection{Performance metrics}

The intraday forecaster produces fully probabilistic outputs, enabling performance evaluation using likelihood-, sample-, quantile-, and point-based metrics. Let the test set be $\mathcal{X}=\{(\mathbf{x}_n,\mathbf{c}_n)\}_{n=1}^N$, where $\mathbf{x}_n=[x_{t,n}]_{t=1}^T = [\mathbf{x}_n^{:T'},\mathbf{x}_n^{T':}]$. For all metrics, lower values indicate better performance.

\subsubsection{Likelihood-based}
Using the closed-form intraday forecast in \eqref{eq:intraday_dist}, we compute the negative loglikelihood (NLL) of the unobserved part $\mathbf{x}_n^{T':}$:
\begin{equation}\label{eq:metric_loglikelihood}
    \text{NLL}(T') = - \frac{1}{N} \sum_n \log \tilde{p}_\psi(\mathbf{x}^{T':}=\mathbf{x}_n^{T':}|\mathbf{x}_n^{:T'}, \mathbf{c}_n).
\end{equation}
Note that $\tilde{p}_\psi(\mathbf{x}|\mathbf{c})$ depends on the sampled CVAE latent variables $\{\mathbf{z}_k\}$, so different samples may yield different NLL values for the same trained CVAE model, $p_\psi(\mathbf{x}|\mathbf{c})$.

\subsubsection{Sample-based}
From the intraday distribution, we form ensembles $\hat{\mathbf{x}}^{T':}_{n,s}\sim\tilde{p}(\mathbf{x}^{T':}|\mathbf{x}_n^{:T'}, \mathbf{c}_n)$ of size $S$.\footnote{In this study, we set $S=K$.} Then we compute the absolute error per time step as
\begin{equation}\label{eq:metric_mae}
    \text{AE}(T',t) = \frac{1}{NS} \sum_{n,s} \lvert x_{t,n}^{T':}-\hat{x}^{T':}_{t,n,s}\rvert,
\end{equation}
with the mean absolute error (MAE) across the remaining horizon
\begin{equation}
    \text{MAE}(T') = \frac{1}{T-T'} \sum_{t=T'+1}^T \text{AE}(T',t).
\end{equation}

\subsubsection{Quantile-based}
We use the same ensembles to form quantile forecasts $y_{t,n,i}^{T':}$ representing the $q_i$-th quantile of the ensemble set $\{ x_{t,n,s}^{T':} \}_{s=1}^S$. Here, $\{q_i\}_{i=1}^Q \subset [0,1]^Q$ is the collection of $Q$ quantile levels with $q_i<q_{i+1}, ~\forall i$. After obtaining the quantile forecasts for each test day, the pinball loss function is used as an approximation of the well-known continuous ranked probability score (CRPS): \cite{fakoor2023flexible}
\begin{equation}
    \zeta(T',t) = \frac{2}{N} \sum_{n,i}  \max( q_i(x_{t,n}^{T':} - y_{t,n,i}^{T':}), (1-q_i)(y_{t,n,i}^{T':} - x_{t,n}^{T':})),
\end{equation}
\begin{equation}\label{eq:metric_cprs}
    \text{CRPS}(T') = \frac{1}{T-T'} \sum_{t=T'+1}^T \zeta(T',t).
\end{equation}

\subsubsection{Point-based}
Lastly, we extract the expected values of the forecasting distributions as the point forecast: $\bar{\mathbf{x}}^{T':}_n = \mathbb{E}_{\tilde{p}_\psi(\mathbf{x}^{T':}|\mathbf{x}^{:T'}_n, \mathbf{c}_n)} \left[ \mathbf{x}^{T':} \right] = \sum_k \gamma_{k,n}^{:T'}  \mu_{k,n}^{T':|:T'}$. We assess the performance of these point forecasts via the root mean squared error (RMSE)
\begin{equation}\label{eq:metric_rmse}
    \text{RMSE}(T') = \frac{1}{N} \sum_{n} \sqrt{ \frac{1}{T-T'} \sum_{t=T'+1}^T \left(x_{t,n}^{T':}-\bar{x}^{T':}_{t,n}\right)^2}.
\end{equation}

\begin{figure*}
    \centering
    \includegraphics[width=1.0\linewidth]{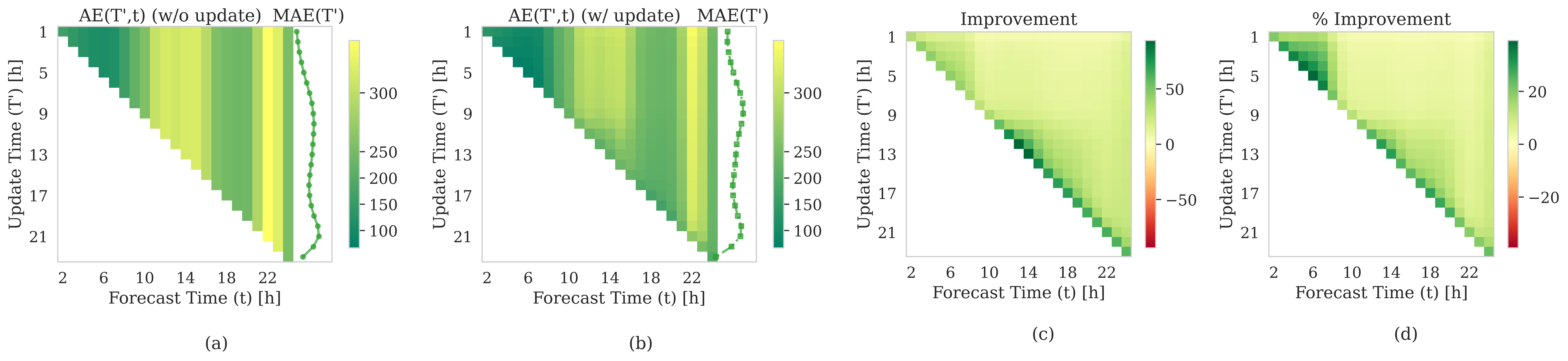}
    \caption{The sample-based metrics AE$(T',t)$ and MAE$(T')$ for the Gipzukoa SM dataset obtained by (a) the baseline non-updated model and (b) the intraday model with the proposed update mechanism. The (c) absolute and (d) percentage differences in the metric AE$(T',t)$ between the models reflect the varying improvements that come with the updates.}
    \label{fig:waterfall_tri}
\end{figure*}

The progressive nature of the updating mechanism within the day makes it cumbersome to summarize the performance with a single number. Instead, we evaluate \textit{performance traces} of the metrics \eqref{eq:metric_loglikelihood}, \eqref{eq:metric_mae}, \eqref{eq:metric_cprs}, and \eqref{eq:metric_rmse} as a function of the update time $T'$. Another way to conceptualize these traces is by considering the single-step performance scores for each combination $(t,T')$ of forecast and update time during the day, as shown in Fig. \ref{fig:waterfall_tri}. For the MAE and CRPS metrics, the performance trace essentially averages these scores over the remaining forecast time.

\subsection{Baselines}

The performance of the proposed intraday forecaster is heavily tied to the modelling accuracy of the day-ahead forecasting distribution. To isolate the effects of the updating mechanism on the day-ahead forecaster, we define a baseline model and a synthetic test set. Note that the purpose of this baseline is to isolate the effect of the Bayesian updating operator while keeping the day-ahead distribution fixed, rather than to compare performance with alternative retraining-based forecasting pipelines.

The baseline model ignores the information provided by the observations at $t=1,\ldots,T'$. It is defined as the non-updated (NU) marginal of the day-ahead distribution starting from time $T'$ up to the end of the day:
\begin{equation}
\begin{split}
    \tilde{p}^{\text{NU}}_\psi(\mathbf{x}^{T':}|\mathbf{x}^{:T'}, \mathbf{c}) :=& ~ \tilde{p}_\psi(\mathbf{x}^{T':}| \mathbf{c}) \\
    =& ~ \frac{1}{K}\sum_k \mathcal{N}(\mathbf{x}_{T':};\mu=\mu_k^{T':}, \Sigma=\Sigma_k^{T':}).
\end{split}
\end{equation}
Note that evaluations using real-world test data also depend on the day-ahead distribution's modelling performance, which can be conflated with the performance of the forecast update process. Thus, we propose a synthetic test set $\mathcal{X}^\text{Synth}=\{(\hat{\mathbf{x}}_n, \mathbf{c}_n)\}_{n=1}^N$ consisting of random samples from the trained model $\hat{\mathbf{x}}_n\sim p_{\psi}(\mathbf{x}_n|\mathbf{c}_n)$. Since it is guaranteed that the sampled data points in this generated test set are ``in-distribution", we ensure that the performance metrics will not be affected by overfitting and the comparison between the two models will reflect only the sample approximation and update effects.

\subsection{Tuning the number of GMM components}
The final crucial step in our experimentation is to determine the appropriate number of GMM components, $K$. Theoretically, using more components results in a better approximation of the CVAE model (see \eqref{eq:cvae_to_gmm}). However, as $K$ increases, so does the computational burden, while the marginal gain diminishes. To find a suitable trade-off, we propose a tuning method that does not rely on any (real) test data.

Our tuning method requires a \textit{best-case} synthetic test set $\mathcal{X}^{\text{Best-case}}=\{(\hat{\mathbf{x}}_n^0, \mathbf{c}_n)\}_{n=1}^N$ that consists of samples from the established day-ahead forecasting distribution in \eqref{eq:cvae_to_gmm}, i.e. $\hat{\mathbf{x}}_n^0 \sim \tilde{p}_{\psi}(\mathbf{x}|\mathbf{c}_n)$. Note the difference to the $\mathcal{X}^\text{Synth}$, which requires a new set of latent samples since the samples must come from the CVAE distribution in \eqref{eq:cvae_to_gmm} that has no closed-form expression. Instead, $\mathcal{X}^{\text{Best-case}}$ utilizes the same set of latent samples as the day-ahead forecaster approximation and guarantees that the test samples are not only in-distribution w.r.t. CVAE, but also the day-ahead GMM forecaster. Intuitively, we supply such samples to the forecaster that the update mechanism can correctly converge to the component $k'$ that the sample of interest came from by assigning a higher weight $\gamma_{k'}^{:T'}$ as time proceeds and observations accumulate.

After obtaining the two test sets for varying $K$, we extract the likelihood-based performance traces for each as $\text{NLL}^{\text{Synth,}K}(T')$ and  $\text{NLL}^{\text{Best-case,}K}(T')$ which are obtained by replacing $\mathbf{x}_n^{T':} (\mathbf{x}_n^{:T'})$ by $\hat{\mathbf{x}}_n^{T':} (\hat{\mathbf{x}}_n^{:T'})$ and ${\hat{\mathbf{x}}_n}^{0^{T':}} ({\hat{\mathbf{x}}_n}^{0^{:T'}})$, respectively. Then, we search for the $K$ that minimizes the average $L_1$ distance between two traces as
\begin{equation}
    K^* = \min_K \frac{1}{T-1} \sum_{T'=1}^{T-1} |\text{NLL}^{\text{Best-case,}K}(T') - \text{NLL}^{\text{Synth,}K}(T')|.
\end{equation}
Intuitively, this optimal $K^*$ results in an approximate distribution $\tilde{p}_{\psi}(\mathbf{x}|\mathbf{c}_n)$ that covers the distribution $p_{\psi}(\mathbf{x}|\mathbf{c}_n)$ well enough that the model performs similarly with regard to forecast updates.

\subsection{Implementation Details}
The CVAE encoder and decoder were implemented as fully connected feed-forward networks with three hidden layers of 1000 neurons each, using ReLU activations. Dropout with probability 0.2 and batch normalisation were applied after each hidden layer. The latent dimensions were set equal to the profile length $L=T$. The pattern dictionary size in the PDCC covariance model is chosen heuristically as $V=100$ for the Gipuzkoa SM dataset and $V=200$ for the Utrecht PV dataset. The models were trained using the Adam optimiser with default parameters in PyTorch 2.4.0 and a batch size of 1024. An early stopping mechanism and a learning rate scheduler were employed based on validation performance. Further implementation details can be found in \cite{bolat2024guide}.
\section{Results and Discussion}

\begin{figure}
\centering
\subfloat[]{\label{fig:num_samples_comp-sm-Loglikelihood}\includegraphics[width=0.95\linewidth]{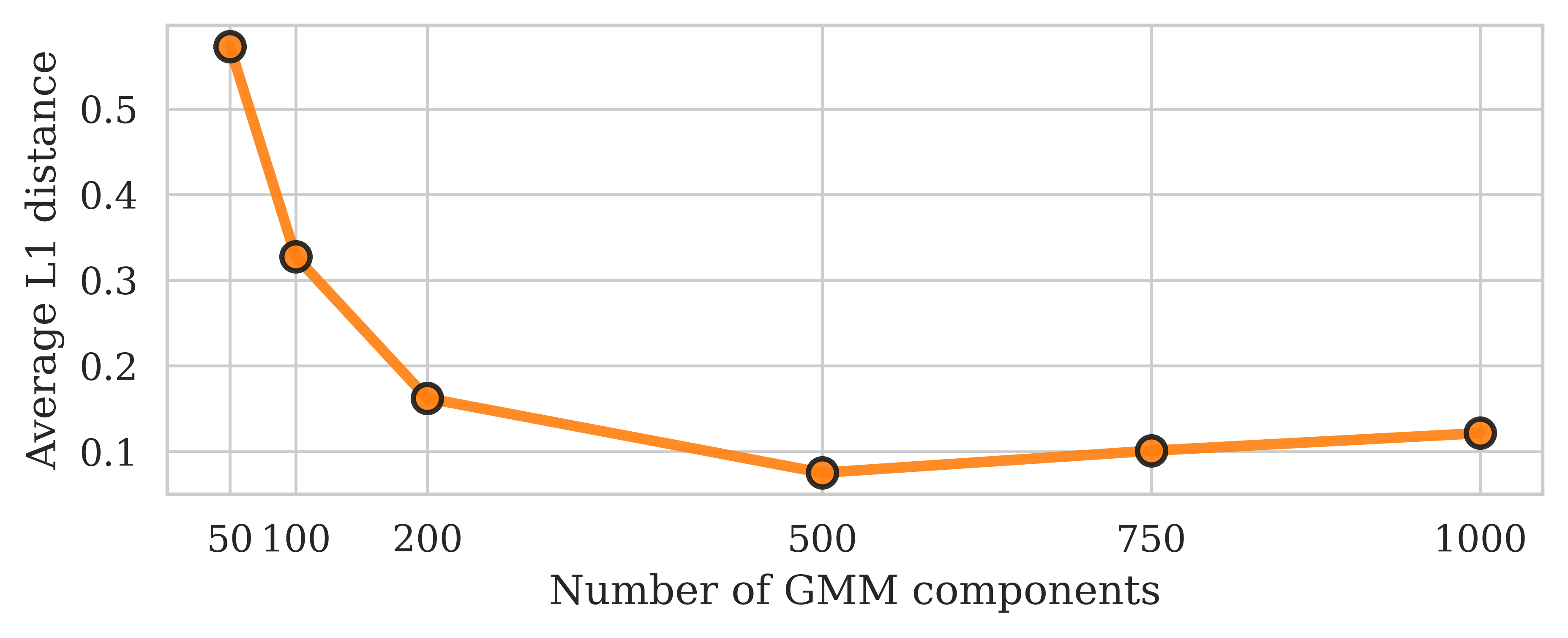}} \\
\subfloat[]{\label{fig:num_samples_comp-pv-Loglikelihood}\includegraphics[width=0.95\linewidth]{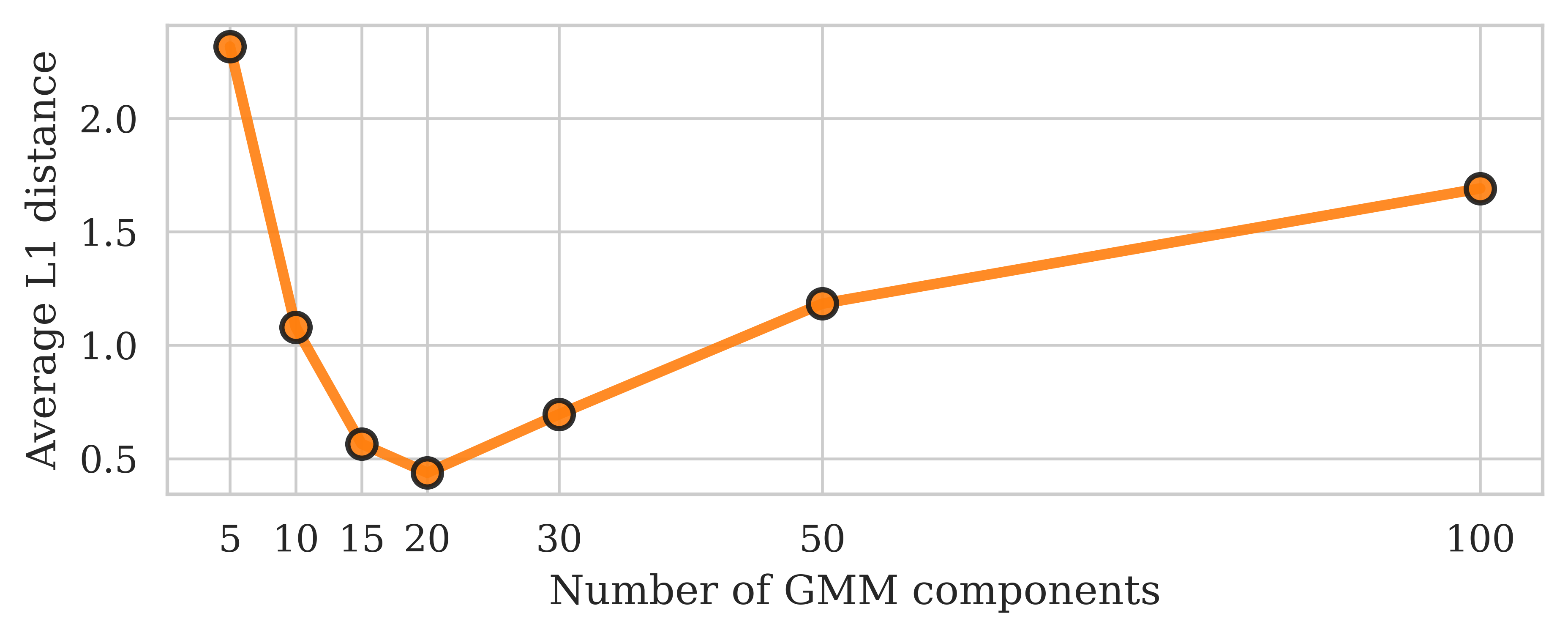}}%
\caption{Tuning of the number of GMM components for the (a) Gipuzkoa SM and (b) Utrecht PV datasets.}
\end{figure}

We begin by presenting the component tuning results for the Gipuzkoa SM and Utrecht PV datasets, shown in Fig.~\ref{fig:num_samples_comp-sm-Loglikelihood} and Fig.~\ref{fig:num_samples_comp-pv-Loglikelihood}, respectively. The tuning procedure suggests employing $K=500$ GMM components for the Gipuzkoa SM dataset and $K=20$ for the Utrecht PV dataset. All subsequent results are based on these settings.

\begin{figure}
    \centering
    \includegraphics[width=1.0\linewidth]{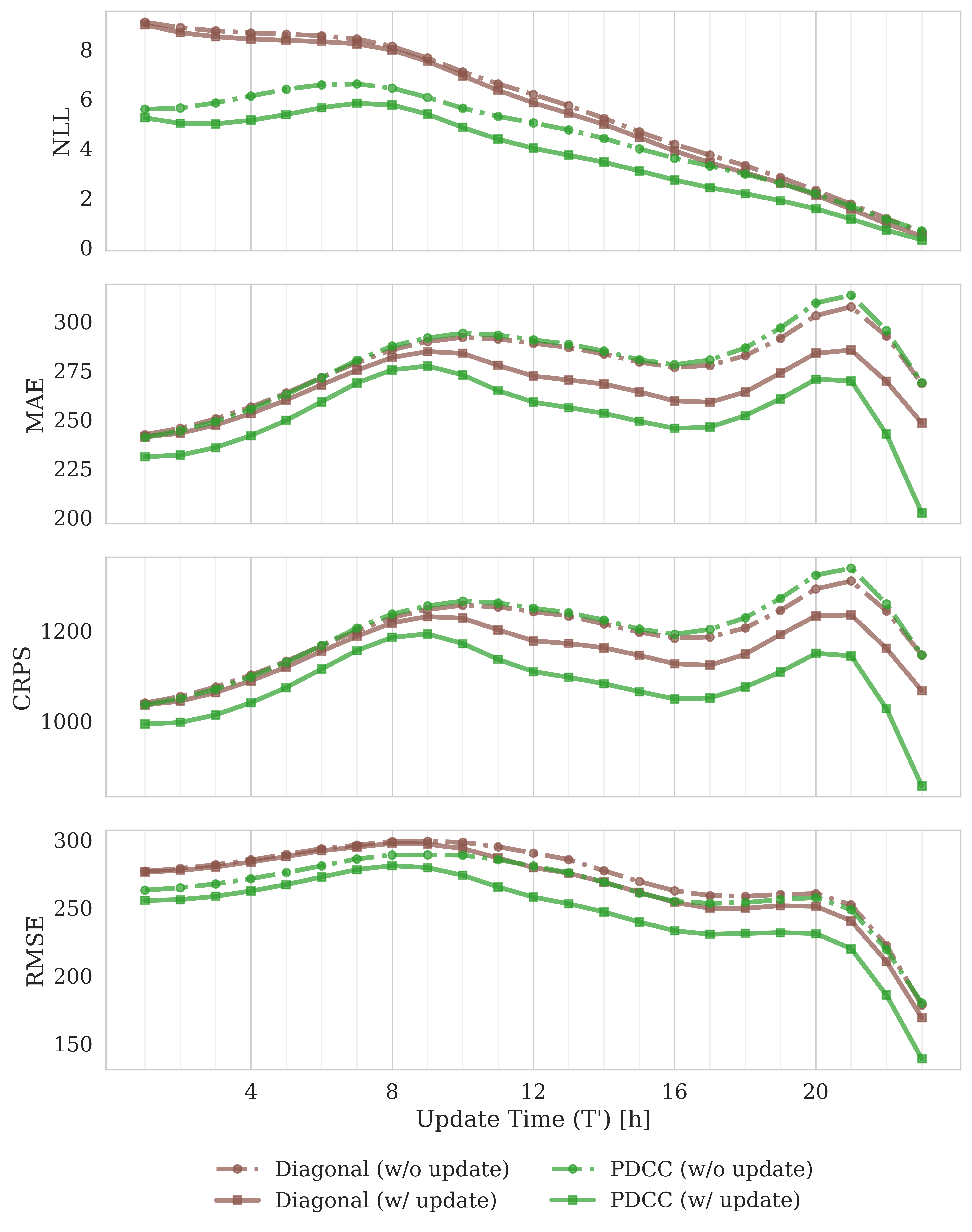}
    \caption{Performance traces over the Gipuzkoa SM synthetic test set. Comparing models with full covariance matrices (PDCC) against diagonal ones.}
    \label{fig:pdcc_vs_diag-sm-all_metrics-synth}
\end{figure}

\begin{figure}
    \centering
    \includegraphics[width=1.0\linewidth]{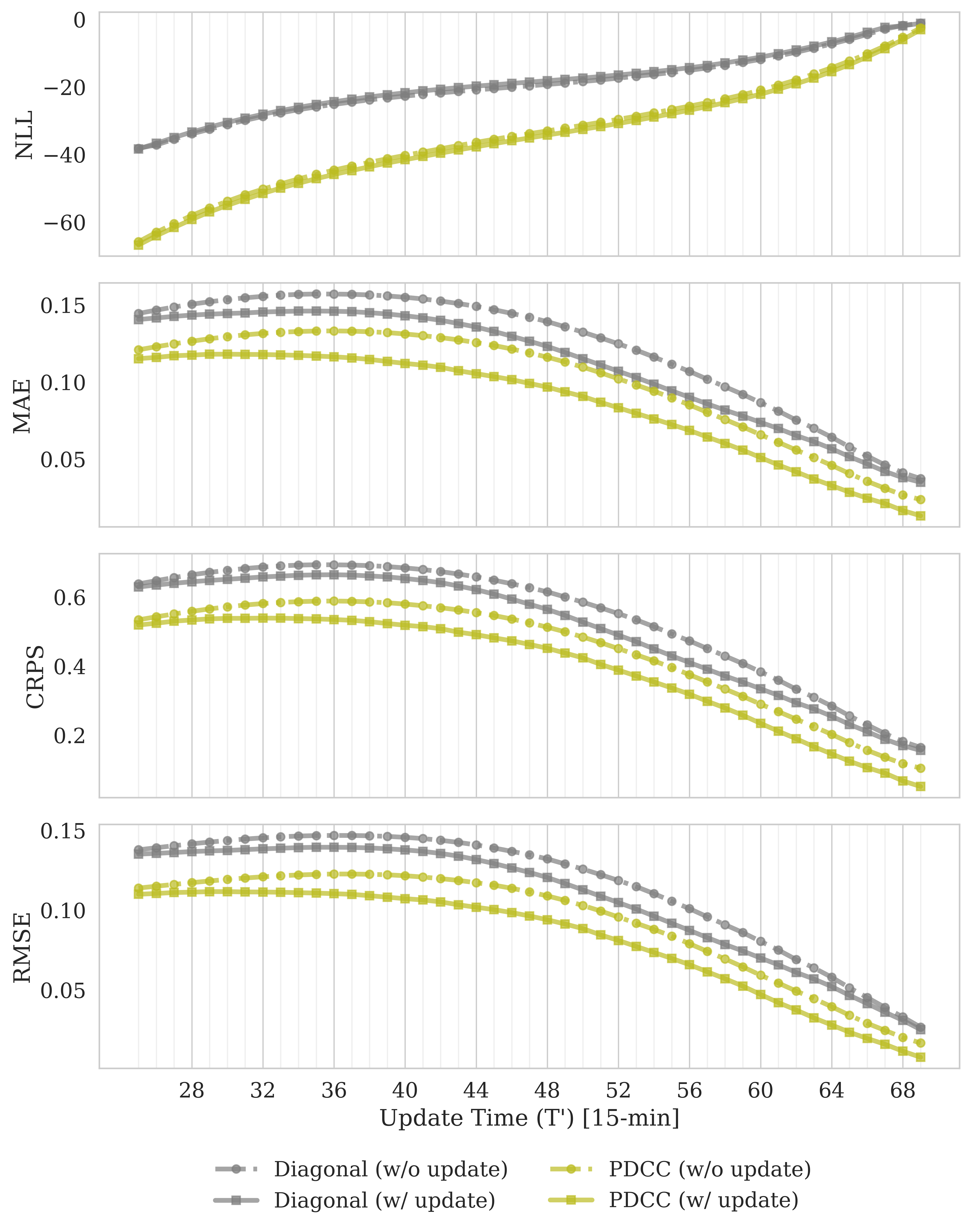}
    \caption{Performance traces over the Utrecht PV synthetic test set. Comparing models with full covariance matrices (PDCC) against diagonal ones.}
    \label{fig:pdcc_vs_diag-pv-all_metrics-synth}
\end{figure}

\begin{figure}
    \centering
    \includegraphics[width=1.0\linewidth]{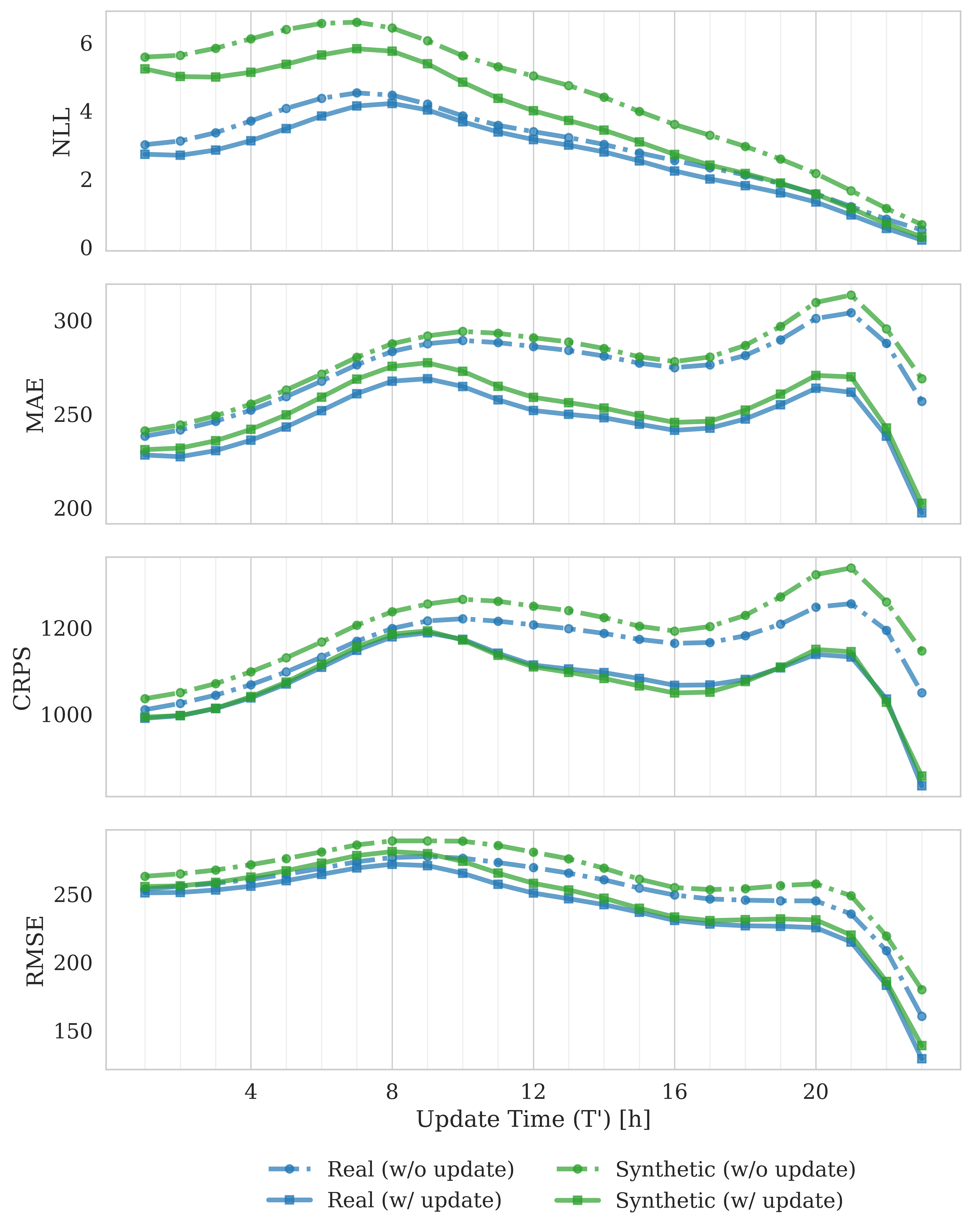}
    \caption{Performance traces of the models with full covariance matrices (PDCC) for the Gipuzkoa SM dataset. Comparing the synthetic test set against the real one.}
    \label{fig:real_vs_synth-sm-all_metrics-pdcc}
\end{figure}

\begin{figure}
    \centering
    \includegraphics[width=1.0\linewidth]{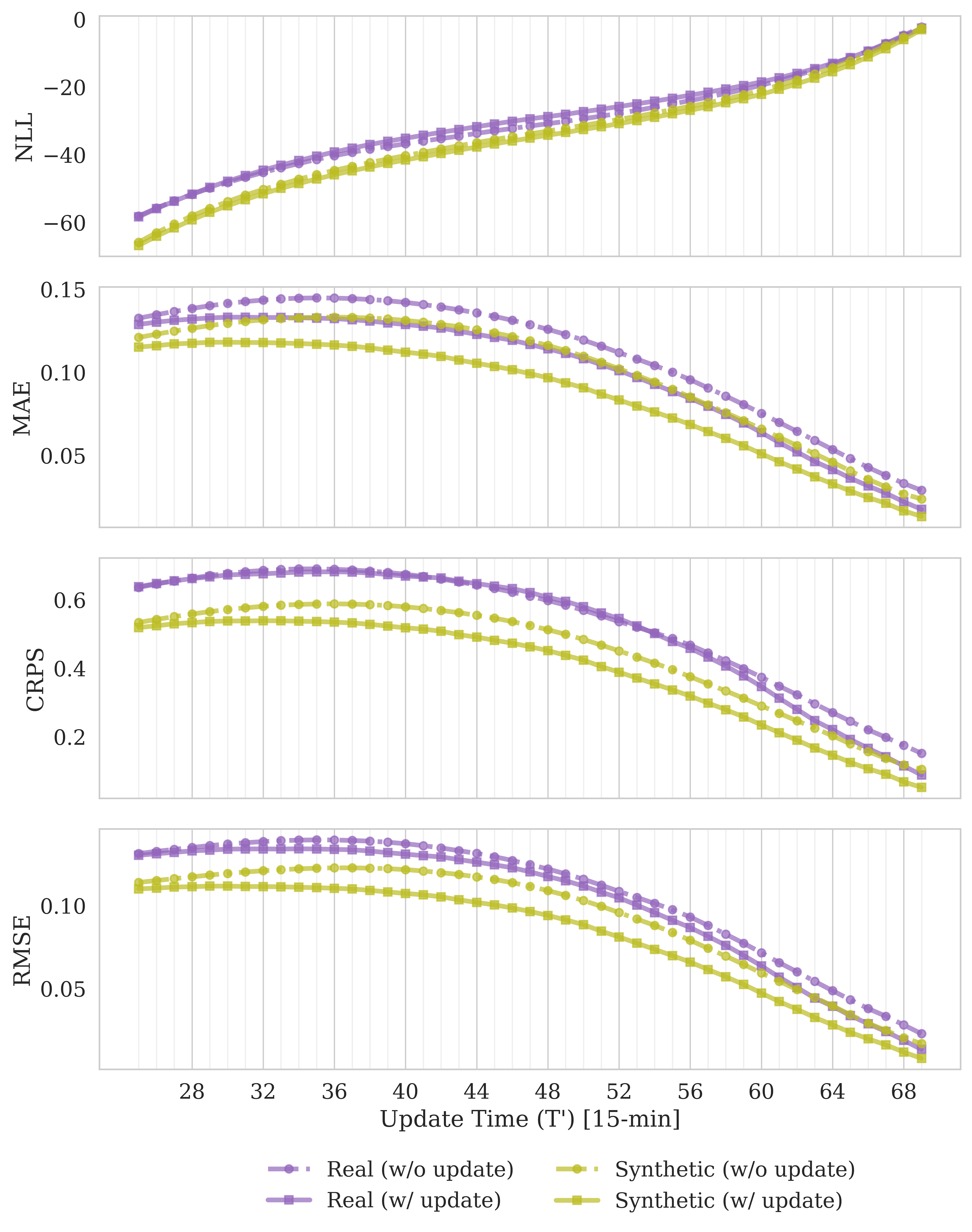}
    \caption{Performance traces of the models with full covariance matrices (PDCC) for the Utrecht PV dataset. Comparing the synthetic test set against the real one.}
    \label{fig:real_vs_synth-pv-all_metrics-pdcc}
\end{figure}

The first general observation from Fig.~\ref{fig:pdcc_vs_diag-sm-all_metrics-synth}--\ref{fig:real_vs_synth-pv-all_metrics-pdcc} is that the Bayesian updating mechanism consistently outperforms the baseline non-updated forecasts across all evaluation metrics. As expected, incorporating observed measurements improves predictive accuracy for the remaining horizon. However, the performance gain is not uniform across update times $T'$. 

This can be examined more closely in Fig.~\ref{fig:waterfall_tri}, which shows that improvements concentrate along the diagonal (near-term improvements), reflecting strong correlations between consecutive time steps. The most recent observations provide the greatest benefit for subsequent predictions, while later hours ($t \gg T'$) are less affected. Localized regions of improvement ($t\in[2,6]$ and $t\in[10,18]$) further indicate that the effectiveness of updates depends also on temporal correlation structures between unobserved steps. Specifically, for the Gipuzkoa SM dataset, the timings of these localized regions are sensible, as the outside regions correspond to the times of leaving and returning home, whereas the preceding measurements generally convey little information regarding those periods. Lastly, note the difference between the absolute and percentage improvements in Fig. \ref{fig:waterfall_tri}(c) and (d), respectively. The discrepancy between them arises from the fact that some regions are already predicted well in the day-ahead. For instance, the $t\in[2,6]$ period for the Gipuzkoa SM dataset is relatively easy to forecast, as household consumption profiles tend to have less uncertainty during early morning periods. Similarly, for the Utrecht PV dataset, Fig. \ref{fig:pdcc_vs_diag-pv-all_metrics-synth} and \ref{fig:real_vs_synth-pv-all_metrics-pdcc} suggest that the biggest improvements stem from updating around noon. This is mainly because early-morning measurements are hardly informative for refining the forecast.

We also observe the effect of employing full covariance matrices for modelling via PDCC against the conventional diagonal structure in Fig. \ref{fig:pdcc_vs_diag-sm-all_metrics-synth} and \ref{fig:pdcc_vs_diag-pv-all_metrics-synth}. For both datasets, it can be seen that PDCC outperforms its diagonal counterpart both in terms of intraday forecast performance (the position of the performance trace with update) and update performance improvement (the difference between the performance traces with and without update). This showcases the higher predictive power that comes with PDCC, as well as its better suitability for the update mechanism, thanks to the explicit modelling of complex temporal correlations. We observed that this relative performance also holds for real data (data not shown).

Finally, we compare performance over real and synthetic test sets in Fig.~\ref{fig:real_vs_synth-sm-all_metrics-pdcc} and Fig.~\ref{fig:real_vs_synth-pv-all_metrics-pdcc}. For the Gipuzkoa SM dataset, performance traces are largely consistent with each other, except for the NLL metric, which suggests that the real test set contains more typical, and therefore easier-to-predict, samples. For the Utrecht PV dataset, on the other hand, the model performs better on the synthetic test set, as expected. Despite this discrepancy, we observe that the update mechanism still provides a performance gain on the real test set in all metrics. The comparatively modest improvement observed for real PV data is likely attributable to the dominant influence of exogenous weather forecast uncertainty and rapid cloud dynamics, which limit the predictive value of early-day measurements for later hours. In addition, the day-ahead model already captures much of the deterministic diurnal solar generation pattern, leaving less room for improvement through conditioning. Consequently, intraday updates yield smaller but still consistent gains across metrics.
\section{Conclusion}

This work proposes a Bayesian updating mechanism to transform day-ahead probabilistic forecasts into intraday forecasts without requiring retraining or re-inference. By conditioning the Gaussian mixture distribution of a CVAE-based day-ahead forecaster on newly observed measurements, the method produces an updated distribution for the remaining hours that preserves its full probabilistic structure. This modelling enables point, quantile, and ensemble forecasts, while being computationally efficient and directly compatible with existing forecasting pipelines.

The experimental results on both electricity consumption and PV generation datasets confirm that the updating mechanism consistently improves forecast accuracy across likelihood-, sample-, quantile-, and point-based metrics. The improvements are most noticeable for time steps with strong temporal correlation to the observed data, including correlations that extend over longer ranges. This highlights the method’s ability to effectively incorporate real-time information. Moreover, the use of pattern dictionary–based covariance structures enhances both the baseline modelling capability and the quality of intraday updates, indicating that intraday conditioning directly benefits from more accurate joint covariance estimation.

Overall, the proposed approach provides a practical and theoretically grounded solution for intraday forecasting in power systems, offering improved adaptability while maintaining probabilistic consistency. Future work will focus on extending the framework with richer exogenous variables, such as enhanced weather forecasts, and exploring its integration into real-time market operations and system balancing tasks.

\bibliographystyle{IEEEtran}
\bibliography{refs.bib} 

\end{document}